\documentclass{article}
\usepackage{spconf,amsmath,graphicx}
\usepackage{cite}

\usepackage{amsmath, amssymb, amsfonts, amsthm}
\usepackage{thmtools}
\usepackage{empheq}
\usepackage{siunitx}
\theoremstyle{definition}

\usepackage{subfloat}
\usepackage{graphicx}
\usepackage[caption=false, font=footnotesize]{subfig}

\usepackage{filecontents}
\usepackage{pgfplots}
\usepackage{tikzscale}
\usepackage{tikz}
\usepgfplotslibrary{patchplots}

\usepackage[switch, pagewise]{lineno}
\usepackage{hyperref}
\hypersetup{
	colorlinks=false,
	hidelinks
}
\usepackage[noabbrev]{cleveref}

\crefname{equation}{Eq.}{Eqs.}
\crefname{figure}{Fig.}{Figs.}
\crefname{algorithm}{Algorithm}{Algorithms}
\crefname{table}{Table.}{Tables.}
\Crefname{table}{Table.}{Tables.}
\crefname{table*}{Table.}{Tables.}
\Crefname{table*}{Table.}{Tables.}

\crefname{lem}{Lemma.}{Lemmas.}
\Crefname{lem}{Lemma.}{Lemmas.}
\crefname{thm}{Theorem.}{Theorems.}
\Crefname{thm}{Theorem.}{Theorems.}
\crefname{prop}{Proposition.}{Propositions.}
\Crefname{prop}{Proposition.}{Propositions.}

\usepackage{multirow}
\usepackage{makecell}

\usepackage{amsthm}

\begin{document}
\topmargin=0mm
% Title.
% ------
\title{SemanticAC: Semantics-assisted
Framework for Audio Classification}
%
% Single address.
% ---------------
\name{Yicheng Xiao$^{1\ast\dagger}$,\quad Yue Ma$^{1\ast}$,
\quad Shuyan Li$^{1}$,
\quad Hantao Zhou$^{1}$,
\quad Ran Liao$^{1}$,
\quad Xiu Li$^{1\ddagger}$\thanks{$^{\ast}$Equal contribution.
$^{\dagger}$Work done during an internship at Tsinghua Shenzhen International Graduate School, Tsinghua University.
$^{\ddagger}$Corresponding author.}
}

\address{\textsuperscript{\rm 1}Tsinghua Shenzhen International Graduate School, Tsinghua University, China
}

\maketitle
\begin{abstract}

In this paper, we propose Semantic\textbf{AC}, a semantics-assisted framework for \textbf{A}udio \textbf{C}lassification to better leverage the semantic information. Unlike conventional audio classification methods that treat class labels as discrete vectors, we employ a language model to extract abundant semantics from labels and optimize the semantic consistency between audio signals and their labels.
We verify that simple textual information from labels and advanced pretraining models enable more abundant semantic supervision for better performance.
Specifically, we design a text encoder to capture the semantic information from the text extension of labels.
Then we map the audio signals to align with the semantics of corresponding class labels via an audio encoder and a similarity calculation module so as to enforce the semantic consistency. 
Extensive experiments on two audio datasets, ESC-50 and US8K demonstrate that our proposed method consistently outperforms the compared audio classification methods.
\end{abstract}
\begin{keywords}
Audio, Classification, Semantics
\end{keywords}
\section{Introduction}
\label{sec:intro}

Audio classification is one of the most essential research subjects in audio deep learning and signal processing.
This type of study can be applied to many practical fields including intelligent transportation\cite{r35}, national security\cite{r36} and healthcare\cite{r37}.
Audio classification is to assign labels to audio signals, which sets the stage for a series of tasks such as 
automatic speech recognition\cite{r38},
keyword spotting\cite{r39},
music genre recognition\cite{r41}, etc.

Over the past decade, most researches in audio classification have emphasized the importance of deep learning.
With the rapid development of convolutional neural network, there have been a lot of great works applying CNN\cite{r7,r8,densenet} to audio event classification.
They employ it to extract features from audio signals and develop a variety of losses to obtain discriminative features.
Inspired by the great encoding power of transformer\cite{r43}, many methods\cite{r10,r11,SEPTR} have been used to model the audio signals with great performance.
Among them, AST\cite{r10} is the first model to introduce self-attention mechanism in audio classification.
HTS-AT\cite{r11} designs a hierarchical transformer structure with great success.
However, applying only a single signal source could not make full use of the abundant information contained in multimedia.
In recent years, multimodal approaches \cite{r28,r16,ma2022simvtp, ma2022visual} have become increasingly popular.
For example, MBT\cite{r28} obtains more discriminative audio feature representations through audiovisual fusion.
AudioCLIP\cite{r16} successfully integrates audio modality into CLIP\cite{r27}, which proves that it is efficient to learn audio representations from visual and natural language supervision.
Nevertheless, the absence of visual signals in most audio datasets motivates us to explore the potential of only using natural language to efficaciously assist audio signal modeling.
\begin{figure}[t]
    \centering
    \includegraphics[width=0.925 \linewidth]{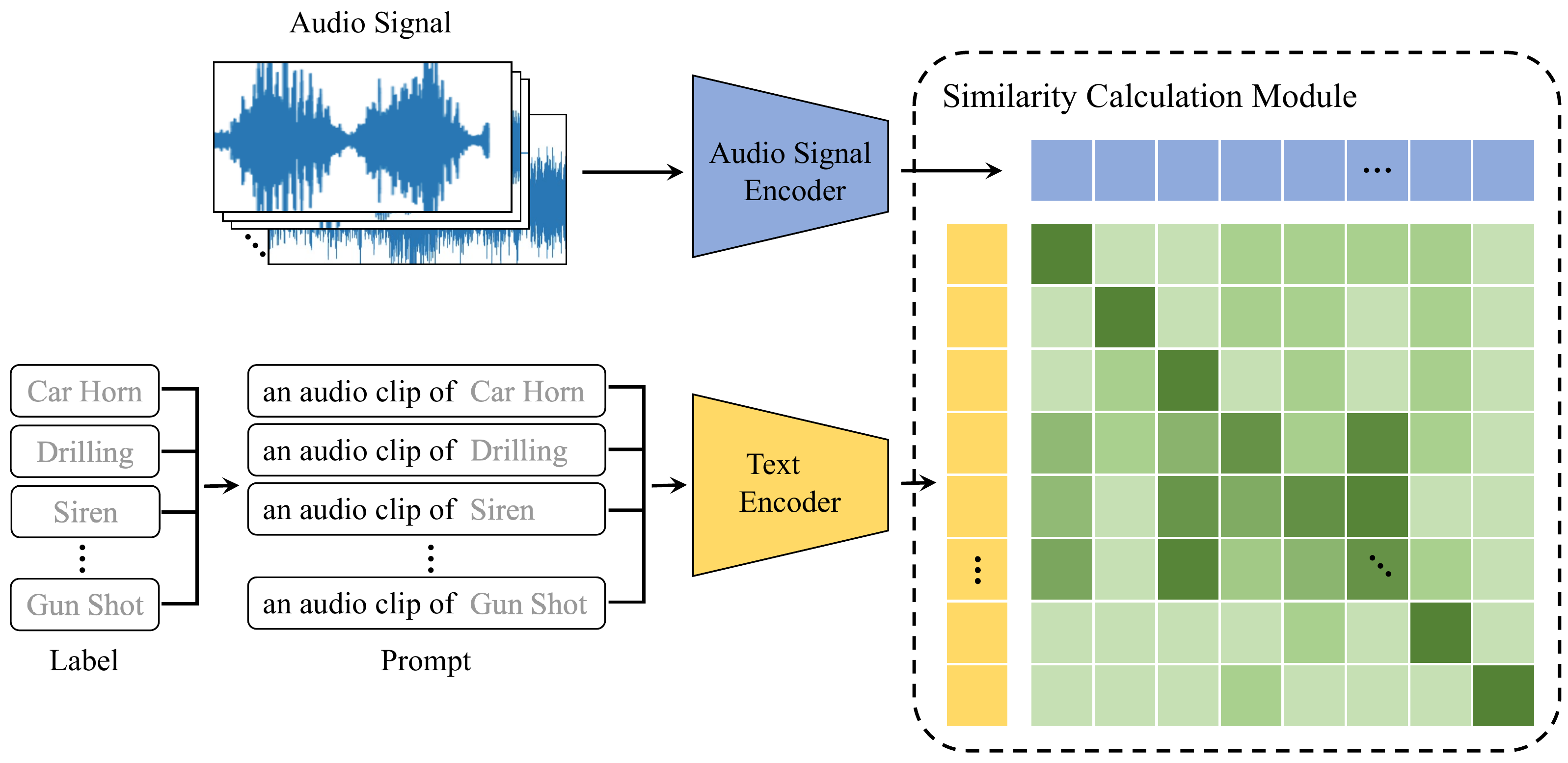}
    \caption{We extract semantics from labels and optimize the semantic consistency between audio signals and their corresponding labels.}
    \label{fig:figure1}
\end{figure}

In this paper, we propose a simple yet effective framework for audio classification, namely SemanticAC.
As shown in \cref{fig:figure1}, the basic idea is to extract the semantic information from classification labels and use it to assist audio modeling.
Specifically, we design a text encoder to map a prompt for label text extension to the semantic representation.
To align the audio feature obtained by a transformer-based structure and the label semantics, we design a lightweight and efficient similarity calculation module and attempt to narrow the gap of these two modalities with contrastive learning.
\begin{figure*}[t]
    \centering
    \includegraphics[width=0.925 \linewidth]{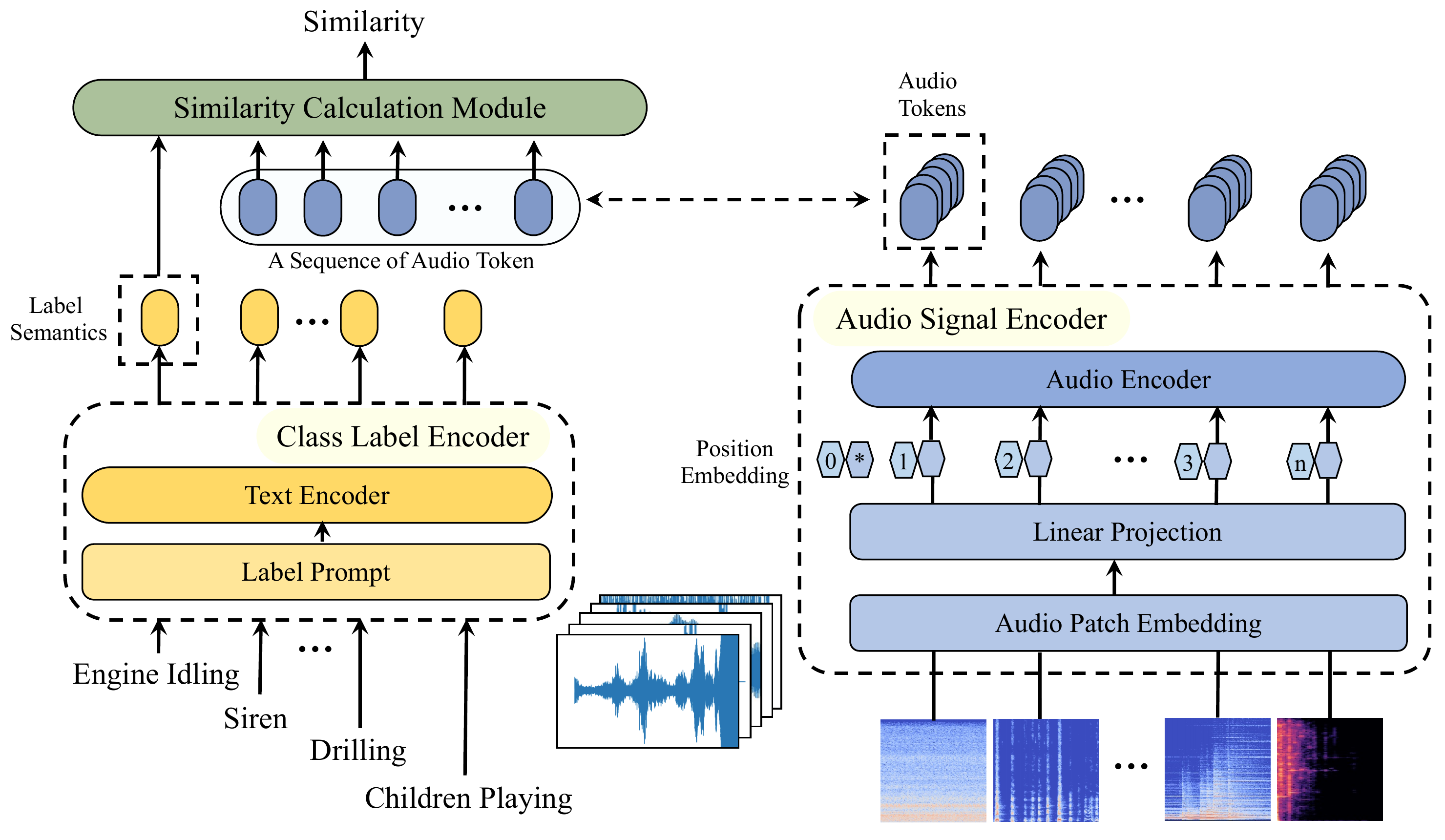}
    \caption{The overview of our proposed semantics-assisted framework for audio classification. }
    \label{fig:figure2}
\end{figure*}

We conduct extensive experiments on two datasets, ESC-50\cite{r2} and US8K\cite{r6} to validate the effectiveness of our proposed SemanticAC.
Our contributions in this paper can be listed as:
\begin{itemize}
    \item We propose an effective semantics-assisted audio-text modalities framework named SemanticAC for audio classification.
    Instead of treating labels as discrete vectors as conventional methods do, our method makes full use of semantic information from labels.
    \item We develop a lightweight and efficient similarity calculation module fully based on CNN structure, namely CSCM to align the audio feature with semantics.
    \item We have achieved consistent and significant improvement on two datasets, ESC-50\cite{r2} and US8K\cite{r6}.
\end{itemize}

We will introduce our framework and experimental validation in detail in the following sections.

\section{SemanticAC}
\label{sec:method}

Given a batch of audio signals, we convert them to mono as one channel by a certain sampling rate.
We then transform them to mel-spectrogram denoted as ${\left\{ x_{i}\right\}} \in R^{F\times T} $, where $F,T$ and $i$ are the dimension of the spectrum feature, the number of time frames and the index of the audio sample, respectively. We aim to map each audio sample ${\left\{ x_{i}\right\}}$ to its class label ${\left\{ y_{i}\right\}} \in Y$:
\begin{equation}
\label{equ: mapping}
    F\left(x_{i}|\theta\right):R^{F\times T}\longmapsto Y
\end{equation}
where $\theta $ is the parameter of the mapping model.

Instead of treating class labels as discrete vectors, i.e. one-hot vectors, our SemanticAC extracts abundant semantics from labels and utilizes the semantic information to assist audio representation learning.
The overview of SemanticAC is illustrated in \cref{fig:figure2}.
We design a text encoder to extract the semantics $\left \{ T_{i}  \right \}\in \Theta ^{C} $ from class labels and an audio encoder to extract audio tokens.
Then we develop a similarity calculation module to get audio feature $\left \{ A_{i}  \right \}\in \Theta ^{C} $ and align it with label semantics in $\Theta ^{C}$, where $\Theta ^{C}$ represents the projected shared embedding space in dimension $C $ such that the audio signal could be mapped to its corresponding class label correctly.

\subsection{Text-Audio Multimodal Encoder}

In this subsection, we describe how to extract the semantics of class labels and audio feature tokens. We design our backbone on the basis of CLIP\cite{r27}, which consists of a class label encoder and an audio signal encoder.

\textbf{Class Label Encoder}.
This encoder is to extract the semantics from label.
Instead of directly injecting the discrete class label to the encoder, we develop a prompt, ``an audio clip of $\left [ LABEL \right ] $", where $LABEL$ represents the label of corresponding audio signal to generate a text extension as input.
This is based on our empirical finding that a simple phrase description can be beneficial to extract semantic information from labels, which will be detailed in \cref{ssec:ablation study}.
In order to extract abundant semantics from label text extension, we design a multi-layer transformer-based text encoder,
which is inspired by the powerful text modeling ability of transformer. 
After text encoding, we get a $C$-dim vector to represent label semantics. 

\textbf{Audio Signal Encoder}.
Firstly, we split the spectrogram of audio signal into patches in order to better capture the correspondence among frequency units of different time frames.
Then we perform a linear layer to project these patches into a sequence.
We replace image-head ViT \cite{r42} of CLIP with a pretrained transformer-based encoder, which utilizes a hierarchical transformer with window attention following HTS-AT\cite{r11} and finally we get a sequence of feature tokens.

\begin{figure}[t]
    \centering
    \includegraphics[width= \linewidth]{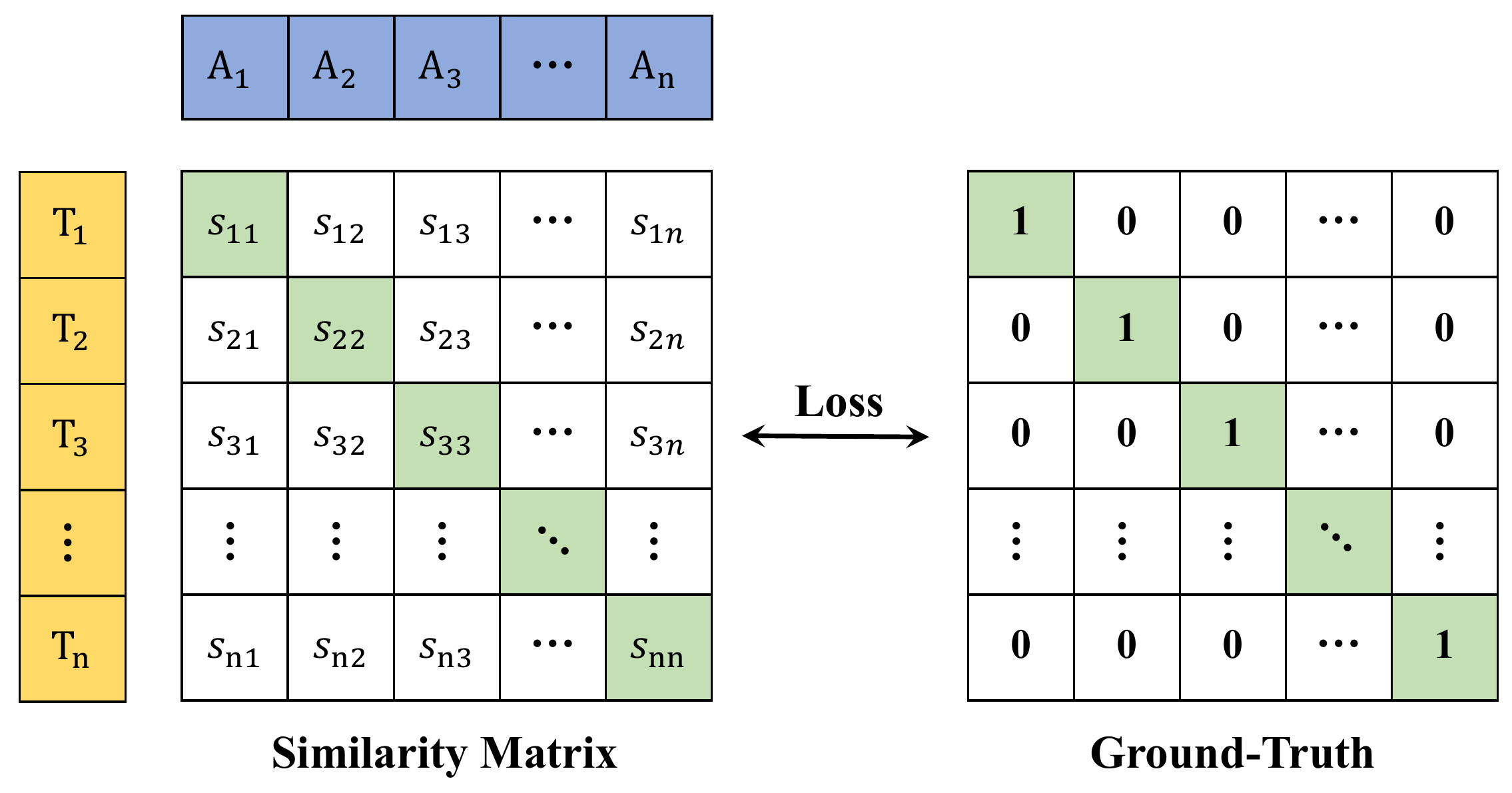}
    \caption{This figure illustrates the similarity matrix $\mathbf{S}_{n\times n} $ of audio and text features, where $s_{ij}=A_{i} \cdot T_{j} \left(i,j=1,2,\cdots,n \right)$ and the corresponding ground-truth.}
    \label{fig:matrix}
\end{figure}

\subsection{Similarity Calculation Module}
\label{ssec:similarity module}
We design a similarity calculation module called CSCM that capture the correlation between the label semantics and audio signal.
Conventional methods directly project a sequence of audio tokens with high dimension to the embedding space with heavy element-wise calculation. In contrast, we employ a series of convolutional networks with proper kernel size to reduce element-wise multiplication, which limits the parameter scale in a small margin.
To generate more abundant feature representation, we apply a convolutional attention mechanism\cite{r34}.
Specifically, it first reshapes the sequence of the audio tokens into a $3D$ feature map $M $ in $\left [ d \times h\times w \right ] $, where $d,h$ and $w$ are depth, height and width of $M$ respectively and then outputs a vector of audio feature with the same dimension $C $ as text representation in the embedding space. 
Finally, we adopt contrastive learning to optimize the whole network.
We measure the distance between two modalities representations via cosine similarity as shown in \cref{equation1}.
\begin{equation}
    \label{equation1}
    s_{ij}=\frac{A_{i}\cdot T_{j}}{\left \|A_{i} \right \|\cdot \left \| T_{j} \right \|  }= \frac{\sum_{m=1}^{C} A^{m}_{i}\times  T^{m}_{j} }{\sqrt{\sum_{m=1}^{C} (A^{m}_{i}})^{2}\times \sqrt{\sum_{m=1}^{C} (T^{m}_{j}})^{2}  } 
\end{equation}
The obtained similarity matrix $\mathbf{S}_{n\times n} $ is shown in \cref{fig:matrix}, where $n $ is the batchsize. %不要引用figure1
The diagonal elements in $\mathbf{S}_{n\times n} $ denote positive sample pairs and all the other elements denote negative sample pairs.
We minimize the distance between positive pairs and maximize the distance between negative pairs via cross entropy loss $CE \left( \right)$ in different axis of matrix. We formulate the loss function as follows, where $\mathbf{S}^{\prime }$ indicates the transpose of $\mathbf{S}$:
\begin{equation}
    \label{equation2}
    Loss = \frac{1}{2}\left [CE\left ( \mathbf{S}, Y  \right ) + CE\left ( \mathbf{S}^{\prime }, Y \right )\right ],
\end{equation}

\section{Experiments}
\label{sec:experiments}

In this section, we evaluate our proposed method on two datasets: ESC-50\cite{r2} and US8K\cite{r6}.

\subsection{Datasets}
\label{ssec:datasets}
The ESC-50\cite{r2} dataset is a collection of label with 2000 short audio clips comprising 50 classes of various environmental sound events arranged into 5 folds. 
The US8K\cite{r6} is an audio dataset that contains 8732 labeled audio excerpts of urban sounds in 10 classes arranged into 10 folds.
For data processing, we resample the signals from ESC-50 into 44100Hz for training and 32000Hz for evaluation. 
As for samples in US8K, we both resample them into 44100Hz for training and evaluation.

\begin{table}[b]
    \centering
    \caption{Evaluation on ESC-50.}
    \resizebox{0.9\linewidth}{!}{
\begin{tabular}{llccc}
\hline
\multicolumn{2}{l}{Model}              & Pretrain           & \begin{tabular}[c]{@{}c@{}}TOP-1  \\         Accuracy\end{tabular} & Average(Acc. \%) \\ \hline
% \multicolumn{2}{l}{MFC\cite{MFC}}           & \checkmark                  & 90.5                                                               &                  \\
\multicolumn{2}{l}{SepTr\cite{SEPTR}}              & -                  & 91.13                                                              & -                  \\
% \multicolumn{2}{l}{ACDNet\cite{ACDNet}}             & -                  & 87.1                                                               & 87.1             \\
% \multicolumn{2}{l}{EAT-S\cite{eat}}                   & \checkmark                  & 95.25                                                              & 95.25            \\
\multicolumn{2}{l}{EAT-M\cite{eat}}              & \checkmark                  & 96.3                                                               & 96.3             \\
\multicolumn{2}{l}{AST\cite{r10}}                & \checkmark                  &  -                                                                  & 95.6             \\
\multicolumn{2}{l}{HTS-AT\cite{r11}}             & \checkmark                  & 97                                                                 & 97               \\
% \multicolumn{2}{l}{L3\cite{l3}}                 & -                  & 79.3                                                               &                  \\
\multicolumn{2}{l}{XDC\cite{xdc}}                & \checkmark                  & 85.4                                                               &    -              \\
\multicolumn{2}{l}{CrissCross\cite{crisscross}}         & \checkmark                  & 90.5                                                               &  -                \\
% \multicolumn{2}{l}{AudioCLIP(partial)\cite{r16}}      & \checkmark                  &                                                                    & 96.65            \\
\multicolumn{2}{l}{AudioCLIP\cite{r16}}         & \checkmark                 &        -                                                            & 97.15            \\
% \multicolumn{2}{l}{AVTS\cite{avts}}               & -                  & 80.6                                                               &                  \\
\multicolumn{2}{l}{AVID\cite{avid}}               & \checkmark                  & 89.2                                                               &    -              \\  \hline
\multicolumn{2}{l}{SemanticAC$^{\ast}$(ours) }         & \multirow{2}{*}{\checkmark} & 96.5                                                               & 96.5             \\
\multicolumn{2}{l}{SemanticAC(ours)} &                    & \textbf{97.25}                                                     & \textbf{97.25}   \\ \hline
\end{tabular}}
\label{tab:esc50}
\end{table}

\subsection{Implementation Details}
\label{ssec:implementation}

The audio encoder 
has been pretrained in AudioSet\cite{r48}.
To fine-tune our semantics-assisted framework, we use SGD\cite{r45} with weight decay of $5e-4$ and batchsize of $16$ to optimize the parameters.
We use ExponentialLR strategy to schedule the learning rate with initialization of $8e-5$, and the decrease actor $\gamma$ is set as $0.96$.
The input text is encoded by the lowercase byte pair of the vocabulary with the size of $49152$ as the same as CLIP.
The dimension $d,h$ and $w$ of $M $ mentioned in \cref{ssec:similarity module}, are set to $768, 8$ and $8$, respectively.
The label semantics and audio feature vector dimension $C $ is set to 1024.
We set the maximum sequence length to $76$ considering the computational resource.
Following \cite{r16}, we apply several data augmentation methods such as Random Crop, Random Noise, etc. to avoid over-fitting.

\begin{table}[t]
    \centering
    \caption{Audio classification accuracy on US8K}
    \resizebox{0.9\linewidth}{!}{
\begin{tabular}{lcc}
\hline
Model              & \begin{tabular}[c]{@{}c@{}}Extra Training\\      Data\end{tabular} & Accuracy (\%) \\ \hline
% Stride-DS-24       & -                                                                  & 70.9             \\
% SB-CNN\cite{sbcnn}             & \checkmark                                                                  & 79               \\
1DCNN\cite{r7}              & \checkmark                                                                  & 89               \\
% EAT-S\cite{eat}              & \checkmark                                                                  & 88.1             \\
DenseNet\cite{densenet}           &                                                                    & 87.42            \\
ESResNeXt\cite{r8}          &                                                                    & 89.14            \\
% Piczak-CNN\cite{piczak}         &                                                                    & 73.7             \\
% AudioCLIP(partial)\cite{r16} & \checkmark                                                                  & 89.95            \\
AudioCLIP\cite{r16}    & \checkmark                                                                  & 90.07            \\
EAT-M\cite{eat}              & \checkmark                                                                  & 90               \\ \hline
SemanticAC(ours)   & \checkmark                                                                  & \textbf{91.34}   \\ \hline
\end{tabular}}
    \label{tab:us8k}
\end{table}

\subsection{Comparison with State-of-the-Art}
\label{ssec:performance}
\begin{figure}[hb]
    \centering
    \includegraphics[width=0.925 \linewidth]{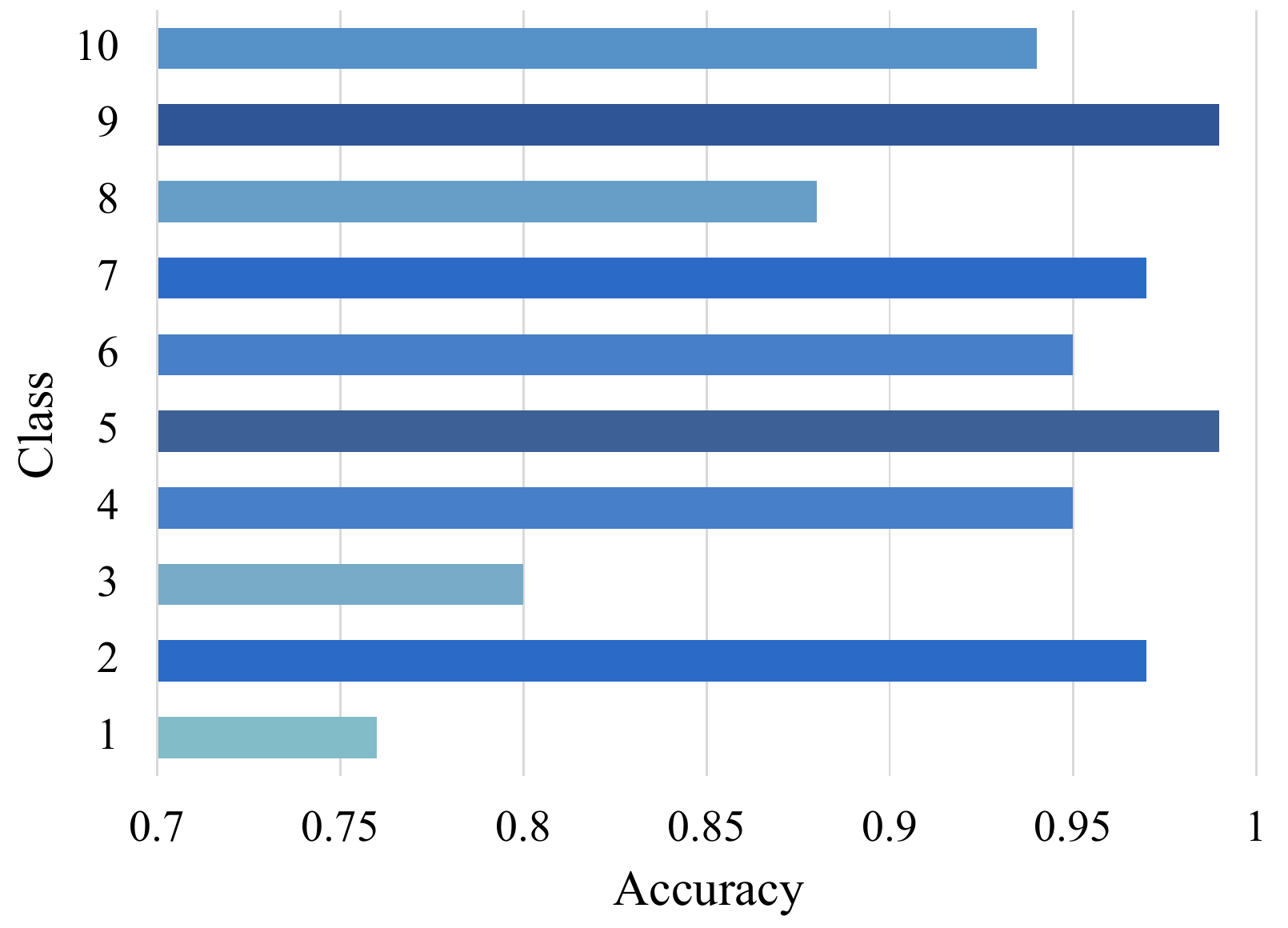}
    % \%hangju{-2mm}
    \caption{The accuracy of $10$ categories on the $8^{th} $ fold of US8K.
    The class number $1-10$ respectively indicates the categories: Air Conditioner, Car Horn, Children Playing, Dog Bark, Drilling, Engine Idling, Gun Shot, Jackhammer, Siren, Street Music.}
    \label{fig:visual}
\end{figure}

We achieve the state-of-the-art(SOTA) $97.25\%$ accuracy on ESC-50 dataset as shown in \cref{tab:esc50}. 
SemanticAC is $0.25\%$ higher than HTS-AT only with audio modality, which indicates the effectiveness of our framework.
We then evaluate on US8K with the same training strategy on ESC-50.
The experiment results in \cref{tab:us8k} show our performance $91.34\%$, which is $1.27\%$ higher than AudioCLIP\cite{r16}, $1.34\%$ higher than EAT-M\cite{eat}.
The visualization in \cref{fig:visual} illustrates that the performance of our method on the $8^{th} $ fold of US8K.

\subsection{Ablation Study}
\label{ssec:ablation study}
We conduct two groups of experiments to validate the effectiveness of each component of our method.

\textbf{Text Assistance and CSCM}.
Our similarity calculation
module aligns the audio feature with the semantic representation from labels through a combination of 
% some convolutional nerual networks,
a convolutional attention mechanism and a CNN-Block,
and finally calculates the cosine similarity.
We employ several convolutional layers with kernel size $3\times3 $ and $1\times 1 $ for feature alignment.
As shown in \cref{tab:module}, SemanticAC$^{\ast}$ represents our method only with text-assistance achieving 96.5\% accuracy on the $4^{th}$ fold of ESC-50, which is 2.3\% higher than baseline and it can achieve $1.5\%$ improvement when we utilize our similarity calculation module.
% where SemanticAC$^{\ast}$ represents our framework without the module CSCM.
We have compared our module with SeqTransf using transformer structure and SeqLSTM using LSTM structure 
% \cite{r46}
mentioned in CLIP4CLIP\cite{r33} by validating on the $4^{th}$ fold of ESC-50.
As depicted in \cref{tab:module}, CSCM can achieve efficient fusion and alignment.
We consider that the text assistance can provide abundant semantic supervision to make audio modeling more effective.
And CSCM is beneficial for narrowing the gap between text and audio.

\textbf{Prompt for Labels}.
We conduct an experiment on ESC-50 as shown in \cref{tab:prompt}.
Compared with using ``$\left [ LABEL \right ] $" as input directly, the prompt, ``a clip of $\left [ LABEL \right ] $" for label text extension can achieve 0.07\% accuracy improvement. 
Meanwhile, it is beneficial for Class Label Encoder to extract abundant significant semantics from labels by further enriching the text input, like ``an audio clip of $\left [ LABEL \right ] $".
It utilizes the fixed collocation to expand the text so as to enhance text modeling.

\begin{table}[t]
\caption{The result of differernt configuration on the $4^{th}$ fold of ESC-50.}
\centerline{
\resizebox{\linewidth}{!}{
\begin{tabular}{c|ccc}
\hline
                                & Text Assistance & Sim. Cal. Module                     & ACC. \% \\ \hline
Baseline                        & -                  & -                             & 94.2             \\
SemanticAC*                     & \checkmark                & -                             & 96.5             \\
\multicolumn{1}{l|}{SemanticAC} & \checkmark                & SeqLSTM   & 97.25            \\
\multicolumn{1}{l|}{SemanticAC} & \checkmark                & SeqTransf & 97.5             \\
\multicolumn{1}{l|}{SemanticAC}                      & \checkmark                & CSCM(ours)                         & \textbf{98}      \\ \hline
\end{tabular}}}
\label{tab:module}
\end{table}

\begin{table}[t]
\caption{The accuracy comparison between different ways of prompt as text input on ESC-50.}
\centerline{
\resizebox{0.8\linewidth}{!}{
\begin{tabular}{c|c}
\hline
Prompt                                       & Average (ACC. \%)                       \\ \hline
$\left [ LABEL \right ]$                     & 97.15(±0.17)                           \\
$a\,clip\,of \left [ LABEL \right ]$         & 97.22(±0.4)                            \\
$an\,audio\,clip\,of \left [ LABEL \right ]$ & 97.25(±0.25) \\ \hline
\end{tabular}}}
\label{tab:prompt}
\end{table}

\section{Conclusion}
\label{sec:conclusion}
In this paper, we present an audio classification framework, namely SemanticAC, which utilizes semantic information from the class labels to assist audio representation learning.
We achieve significant improvements on two audio datasets, ESC-50 and US8K.
In the future, we will explore using more sources of signals to assist audio classification.
% Below is an example of how to insert images. Delete the ``\%%vspace'' line,
% uncomment the preceding line ``\centerline...'' and replace ``imageX.ps''
% with a suitable PostScript file name.
% -------------------------------------------------------------------------
% \begin{figure}[htb]

% \begin{minipage}[b]{1.0\linewidth}
%   \centering
%   \centerline{\includegraphics[width=8.5cm]{./image1.eps}}
% %  \%%vspace{2.0cm}
%   \centerline{(a) Result 1}\medskip
% \end{minipage}
% %
% \begin{minipage}[b]{.48\linewidth}
%   \centering
%   \centerline{\includegraphics[width=4.0cm]{./image3.eps}}
% %  \%%vspace{1.5cm}
%   \centerline{(b) Results 3}\medskip
% \end{minipage}
% \hfill
% \begin{minipage}[b]{0.48\linewidth}
%   \centering
%   \centerline{\includegraphics[width=4.0cm]{./image4.eps}}
% %  \%%vspace{1.5cm}
%   \centerline{(c) Result 4}\medskip
% \end{minipage}
% %
% \caption{Example of placing a figure with experimental results.}
% \label{fig:res}
% %
% \end{figure}

%\vfill
%\pagebreak

% \vfill\pagebreak

\bibliographystyle{IEEEbib}
\bibliography{refs, strings}

\begin{thebibliography}{10}

\bibitem{r35}
D.~Williams, D.~De~Martini, M.~Gadd, L.~Marchegiani, and P.~Newman,
\newblock ``Keep off the grass: Permissible driving routes from radar with weak
  audio supervision,''
\newblock in {\em ITSC}. IEEE, 2020, pp. 1--6.

\bibitem{r36}
S.~Shan, J.~Liu, and Y.~Dun,
\newblock ``Prospect of voiceprint recognition based on deep learning,''
\newblock in {\em JPCS}. IOP Publishing, 2021, vol. 1848, p. 012046.

\bibitem{r37}
M.~Esposito, G.~Uehara, and A.~Spanias,
\newblock ``Quantum machine learning for audio classification with applications
  to healthcare,''
\newblock in {\em IISA}. IEEE, 2022, pp. 1--4.

\bibitem{r38}
C.~Zoril{\u{a}} and R.~Doddipatla,
\newblock ``Speaker reinforcement using target source extraction for robust
  automatic speech recognition,''
\newblock in {\em ICASSP}. IEEE, 2022, pp. 6297--6301.

\bibitem{r39}
G.~Chen, C.~Parada, and G.~Heigold,
\newblock ``Small-footprint keyword spotting using deep neural networks,''
\newblock in {\em ICASSP}. IEEE, 2014, pp. 4087--4091.

\bibitem{r41}
D.~Ghosal and M.~H. Kolekar,
\newblock ``Music genre recognition using deep neural networks and transfer
  learning.,''
\newblock in {\em Interspeech}, 2018, pp. 2087--2091.

\bibitem{r7}
S.~Abdoli, P.~Cardinal, and A.~L. Koerich,
\newblock ``End-to-end environmental sound classification using a 1d
  convolutional neural network,''
\newblock {\em ESWA}, vol. 136, pp. 252--263, 2019.

\bibitem{r8}
A.~Guzhov, F.~Raue, J.~Hees, and A.~Dengel,
\newblock ``Esresne (x) t-fbsp: Learning robust time-frequency transformation
  of audio,''
\newblock in {\em IJCNN}. IEEE, 2021, pp. 1--8.

\bibitem{densenet}
K.~Palanisamy, D.~Singhania, and A.~Yao,
\newblock ``Rethinking cnn models for audio classification,''
\newblock {\em arXiv preprint arXiv:2007.11154}, 2020.

\bibitem{r43}
A.~Vaswani, N.~Shazeer, N.~Parmar, J.~Uszkoreit, L.~Jones, A.~N. Gomez,
  {\L}.~Kaiser, and I.~Polosukhin,
\newblock ``Attention is all you need,''
\newblock {\em NeurIPS}, vol. 30, 2017.

\bibitem{r10}
Y.~Gong, Y.-A. Chung, and J.~Glass,
\newblock ``Ast: Audio spectrogram transformer,''
\newblock {\em arXiv preprint arXiv:2104.01778}, 2021.

\bibitem{r11}
K.~Chen, X.~Du, B.~Zhu, Z.~Ma, T.~Berg-Kirkpatrick, and S.~Dubnov,
\newblock ``Hts-at: A hierarchical token-semantic audio transformer for sound
  classification and detection,''
\newblock in {\em ICASSP}. IEEE, 2022, pp. 646--650.

\bibitem{SEPTR}
N.-C. Ristea, R.~T. Ionescu, and F.~S. Khan,
\newblock ``Septr: Separable transformer for audio spectrogram processing,''
\newblock {\em arXiv preprint arXiv:2203.09581}, 2022.

\bibitem{r28}
A.~Nagrani, S.~Yang, A.~Arnab, A.~Jansen, C.~Schmid, and C.~Sun,
\newblock ``Attention bottlenecks for multimodal fusion,''
\newblock {\em NeurIPS}, vol. 34, pp. 14200--14213, 2021.

\bibitem{r16}
A.~Guzhov, F.~Raue, J.~Hees, and A.~Dengel,
\newblock ``Audioclip: Extending clip to image, text and audio,''
\newblock in {\em ICASSP}. IEEE, 2022, pp. 976--980.

\bibitem{ma2022simvtp}
Y.~Ma, T.~Yang, Y.~Shan, and X.~Li,
\newblock ``Simvtp: Simple video text pre-training with masked autoencoders,''
\newblock {\em arXiv preprint arXiv:2212.03490}, 2022.

\bibitem{ma2022visual}
Y.~Ma, Y.~Wang, Y.~Wu, Z.~Lyu, S.~Chen, X.~Li, and Y.~Qiao,
\newblock ``Visual knowledge graph for human action reasoning in videos,''
\newblock in {\em Proceedings of the 30th ACM International Conference on
  Multimedia}, 2022, pp. 4132--4141.

\bibitem{r27}
A.~Radford, J.~W. Kim, C.~Hallacy, A.~Ramesh, G.~Goh, S.~Agarwal, G.~Sastry,
  A.~Askell, P.~Mishkin, J.~Clark, et~al.,
\newblock ``Learning transferable visual models from natural language
  supervision,''
\newblock in {\em ICML}. PMLR, 2021, pp. 8748--8763.

\bibitem{r2}
K.~J. Piczak,
\newblock ``Esc: Dataset for environmental sound classification,''
\newblock in {\em ACM MM}, 2015, pp. 1015--1018.

\bibitem{r6}
J.~Salamon, C.~Jacoby, and J.~P. Bello,
\newblock ``A dataset and taxonomy for urban sound research,''
\newblock in {\em ACM MM}, 2014, pp. 1041--1044.

\bibitem{r42}
A.~Dosovitskiy, L.~Beyer, A.~Kolesnikov, D.~Weissenborn, X.~Zhai,
  T.~Unterthiner, et~al.,
\newblock ``An image is worth 16x16 words: Transformers for image recognition
  at scale,''
\newblock {\em arXiv preprint arXiv:2010.11929}, 2020.

\bibitem{r34}
S.~Woo, J.~Park, J.-Y. Lee, and I.~S. Kweon,
\newblock ``Cbam: Convolutional block attention module,''
\newblock in {\em ECCV}, 2018, pp. 3--19.

\bibitem{eat}
A.~Gazneli, G.~Zimerman, T.~Ridnik, G.~Sharir, and A.~Noy,
\newblock ``End-to-end audio strikes back: Boosting augmentations towards an
  efficient audio classification network,''
\newblock {\em arXiv preprint arXiv:2204.11479}, 2022.

\bibitem{xdc}
H.~Alwassel, D.~Mahajan, B.~Korbar, L.~Torresani, B.~Ghanem, and D.~Tran,
\newblock ``Self-supervised learning by cross-modal audio-video clustering,''
\newblock {\em NeurIPS}, vol. 33, pp. 9758--9770, 2020.

\bibitem{crisscross}
P.~Sarkar and A.~Etemad,
\newblock ``Self-supervised audio-visual representation learning with relaxed
  cross-modal temporal synchronicity,''
\newblock {\em arXiv preprint arXiv:2111.05329}, 2021.

\bibitem{avid}
P.~Morgado, N.~Vasconcelos, and I.~Misra,
\newblock ``Audio-visual instance discrimination with cross-modal agreement,''
\newblock in {\em CVPR}, 2021, pp. 12475--12486.

\bibitem{r48}
J.~F. Gemmeke, D.~P. Ellis, D.~Freedman, et~al.,
\newblock ``Audio set: An ontology and human-labeled dataset for audio
  events,''
\newblock in {\em ICASSP}. IEEE, 2017, pp. 776--780.

\bibitem{r45}
B.~T. Polyak and A.~B. Juditsky,
\newblock ``Acceleration of stochastic approximation by averaging,''
\newblock {\em SIAM J CONTROL OPTIM}, vol. 30, no. 4, pp. 838--855, 1992.

\bibitem{r33}
H.~Luo, L.~Ji, M.~Zhong, Y.~Chen, W.~Lei, N.~Duan, and T.~Li,
\newblock ``Clip4clip: An empirical study of clip for end to end video clip
  retrieval and captioning,''
\newblock {\em Neurocomputing}, vol. 508, pp. 293--304, 2022.

\end{thebibliography}

\end{document}